\documentclass[12pt]{article}
\usepackage[latin9]{inputenc}
\usepackage{geometry}
\usepackage{units}
\usepackage{bm}
\usepackage{amsmath}
\usepackage{amssymb}
\usepackage{cancel}
\usepackage{graphicx}

\makeatletter

\providecommand{\tabularnewline}{\\}

\global\long\def\order#1{\mathcal{O}\left(#1\right)}
\global\long\def\d{\mathrm{d}}

\def\za{Z\alpha}
\def\order#1{{\cal O}\left(#1\right)}
\def\az{\alpha_Z}

\usepackage{slashed}
\usepackage{bm}

\makeatother

\begin{document}
\global\long\def\gZ{\Gamma\left[ \left( Z\mu^{-} \right) \rightarrow\left( Ze^{-} \right) \nu_{\mu}\bar{\nu}_{e} \right]}%
\global\long\def\tPhi{\widetilde{\Phi}(\bm{k})}%
\global\long\def\tf{\widetilde{f}\left(k\right)}%
\global\long\def\tg{\widetilde{g}\left(k\right)}%
\global\long\def\calM{\mathcal{M}}%
\global\long\def\vk{\bm{k}}%
\global\long\def\vn{\bm{n}}%
\global\long\def\vp{\bm{p}}%
\global\long\def\vq{\bm{q}}%
\global\long\def\vr{\bm{r}}%
\global\long\def\vs{\bm{s}}%
\global\long\def\vu{\bm{u}}%
\global\long\def\vv{\bm{v}}%
\global\long\def\vw{\bm{w}}%
\global\long\def\vx{\bm{x}}%
\global\long\def\vy{\bm{y}}%
\global\long\def\vz{\bm{z}}%
\global\long\def\vrho{\bm{\rho}}%
\global\long\def\vsi{\bm{\sigma}}%

\global\long\def\order#1{\mathcal{O}\left(#1\right)}%

\global\long\def\d{\mbox{d}}%

\global\long\def\tr{\mbox{Tr}}%

\global\long\def\Li{\mbox{Li}_{2}}%

\global\long\def\az{\alpha_{Z}}%

\global\long\def\za{Z\alpha}%

\global\long\def\Eb#1{E_{{\scriptscriptstyle \text{bind},#1}}}%

\global\long\def\GF{G_{{\scriptscriptstyle \text{F}}}}%

\title{Decay of a bound muon into a bound electron}
\author{M. Jamil Aslam$^{\text{(a,b)}}$, Andrzej
  Czarnecki$^{\text{(a)}}$, Guangpeng Zhang$^{\text{(a)}}$, and
  Anna Morozova$^{\text{(a)}}$\\
\textit{\small{}$^{\text{(a)}}$Department of Physics, University
of Alberta, Edmonton, Alberta, Canada T6G 2E1}\\
\textit{\small{}$^{\text{(b)}}$Department of Physics, Quaid-i-Azam
University, Islamabad, Pakistan}}
\date{\vspace*{-60mm}\hfill Alberta Thy 15-19 \vspace*{60mm}}
\maketitle
\begin{abstract}
When a muon bound in an atom decays, there is a small probability
that the daughter electron remains bound. That probability is evaluated.
Surprisingly, a significant part of the rate is contributed by the
negative energy component of the wave function, neglected in a previous
study. A simple integral representation of the rate is presented.
In the limit of close muon and electron masses, an analytic formula
is derived. 
\end{abstract}

\section{Introduction}

Electrostatic binding of a muon $\mu^{-}$ in an atom changes its
decay characteristics. Coulomb attraction decreases the phase space
available to the decay products but enhances the daughter electron
wave function. Muon motion smears the energy spectrum of electrons.
All these effects largely cancel in the lifetime of the muon \cite{Uberall:1960zz}
but they do slow down the decay by a factor that, for small atomic
numbers $Z$, reads 
\begin{equation}
\Gamma\left(\left(Z\mu^{-}\right)\to e\nu_{\mu}\overline{\nu}_{e}N\right)=\left(1-\frac{\left(Z\alpha\right)^{2}}{2}\right)\Gamma\left(\mu\to e\nu_{\mu}\overline{\nu}_{e}\right),\label{eq:TimeDil}
\end{equation}
and can be interpreted as the time dilation; the characteristic velocity
of the bound muon is $Z\alpha$. 

Another possible effect, of primary interest in this paper, is the
decay into an electron that remains bound to the nucleus $N$. For
the actual small ratio of electron to muon masses, $\nicefrac{m_{e}}{m_{\mu}}\simeq\nicefrac{1}{207}$,
that process is very rare, especially for weak binding in atoms with
moderate $Z$. We study it as a part of a program of characterizing
bound muon decays, motivated by upcoming experiments COMET \cite{Shoukavy:2019ydh}
and Mu2e \cite{Bernstein:2019fyh}.

Throughout this paper we use $c=\hbar=1$ and treat the nucleus $N$
as static, spin 0, and point-like, neglecting effects of its recoil
and finite size. We denote its number of protons by $Z$. The notation
$\left(Z\mu^{-}\right)$ or $\left(Ze^{-}\right)$ denotes a muon
or an electron bound in the 1s state, forming a hydrogen-like atom.
We assume that no other particles are bound to the nucleus (we neglect
screening or Pauli blocking due to other electrons).

The decay $\left(Z\mu^{-}\right)\to\left(Ze^{-}\right)\nu_{\mu}\overline{\nu}_{e}$
was previously studied in the very elegant and detailed paper \cite{Greub:1994fp}.
We reevaluate it and find discrepancies with that pioneering study,
particularly for large values of $Z$. This is most likely explained
by negative energy components of the Dirac wave functions, neglected
in \cite{Greub:1994fp} (as discussed in its Appendix A). Here we
use exact Dirac wave functions in the Coulomb field of a point-like
nucleus. Effects of extended nuclear charge distribution were found
to be very small in \cite{Greub:1994fp} so we neglect them. 

Earlier studies of the differences between the decay of a free and
of a bound muon include \cite{Porter:1951zz,gilinsky60,huff61,Haenggi:1974hp}.
More recently, the spectrum of produced electrons was determined in
\cite{Czarnecki:2014cxa,Szafron:2016cbv,Czarnecki:2011mx}. 

This paper is organized as follows. In Section \ref{sec:k-space},
momentum space wave functions are used to compute the rate $\Gamma\left[\left(Z\mu^{-}\right)\to\left(Ze^{-}\right)\nu_{\mu}\overline{\nu}_{e}\right]$,
as in Ref.~\cite{Greub:1994fp}. Significant differences are found
so the result is checked with position space wave functions in Section
\ref{sec:R-space}. That approach turns out to be much simpler; a
one-dimensional integral representation is found, replacing the triple
integral of Ref.~\cite{Greub:1994fp}. In the limit of nearly equal
masses, $m_{e}\to m_{\mu}$, the remaining integration is done analytically
and a closed formula for the rate is obtained in Section \ref{sec:EqualMasses}.
Section \ref{sec:Conclusion} presents conclusions and Appendix \ref{sec:Alchemy}
summarizes the formalism of Ref.~\cite{Greub:1994fp}.

\section{Momentum space derivation of the decay rate \label{sec:k-space}}

\subsection{Wave function and its normalization in momentum space}

\begin{figure}
\centering\includegraphics[scale=0.9]{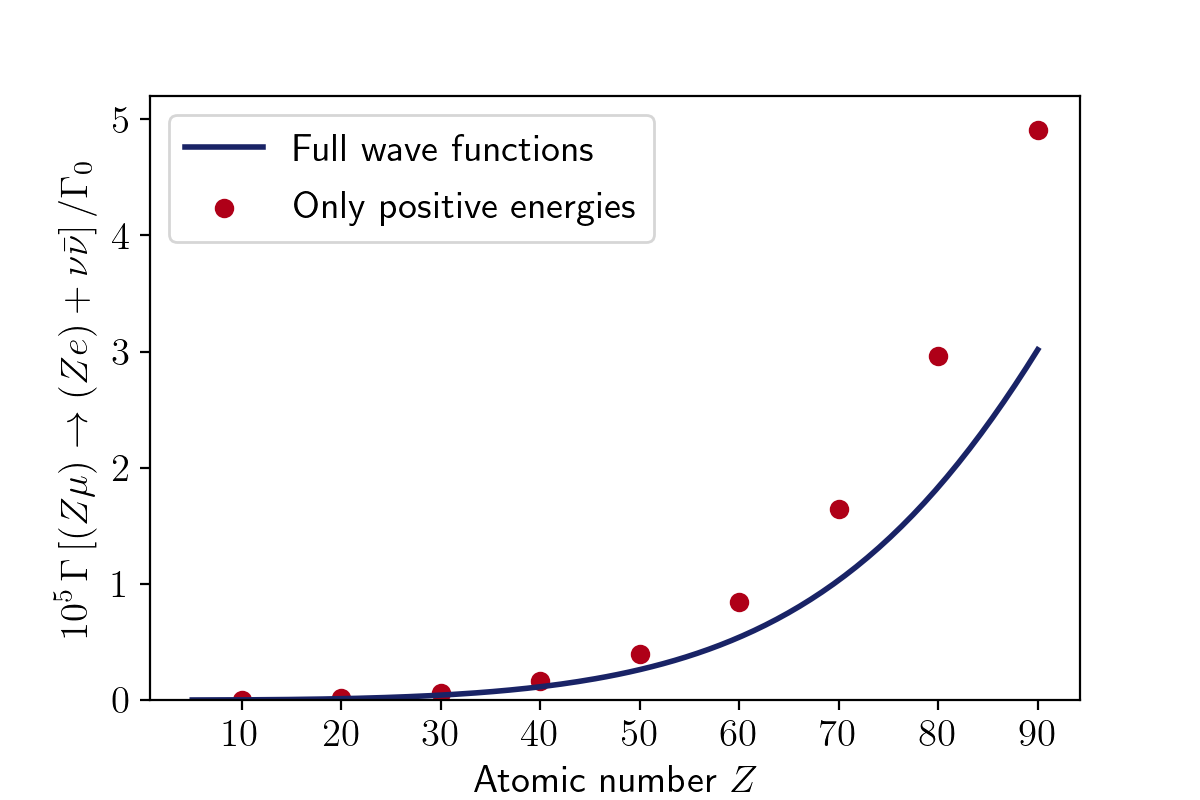}

\caption{Rate of the bound-to-bound decay $\left(Z\mu^{-}\right)\to\left(Ze^{-}\right)+\nu\bar{\nu}$
normalized to the free muon decay rate $\Gamma_{0}$, as a function
of the atomic number $Z$. The solid line shows our results found
using complete wave functions. Red dots are the values obtained by
neglecting negative energy components of wave functions. Errors due
to this approximation grow with the atomic number $Z$. \label{fig:Rate-as-fnc-of-Z}}

\end{figure}

We consider the muon in the ground state of a hydrogen-like ion and
are interested in the final-state electron also in the ground state.
Both muon and electron wave functions are 1s solutions of the Dirac
equation and differ only by the mass, respectively $m_{\mu}$ and
$m_{e}$. Below we present formulas for a generic mass $m$. The position
space wave function $\Phi\left(\vx\right)$ will be presented below
in eq. \eqref{eq:O1}. Taking its Fourier transform we obtain
\begin{align}
\widetilde{\Phi}_{\pm}\left(\vk\right) & =\int\d^{3}x\Phi\left(\vx\right)e^{-i\vk\cdot\vx}\label{Alch:A1-1}\\
 & =\left(\begin{array}{c}
\tf\phi_{\pm}\\
\tg\frac{\vsi\cdot\vk}{k}\phi_{\pm}
\end{array}\right)\quad k=\left|\vk\right|,\label{Alch:A2-1}
\end{align}
where $\phi_{+}=\left(\begin{array}{c}
1\\
0
\end{array}\right)$ and $\phi_{-}=\left(\begin{array}{c}
0\\
1
\end{array}\right)$ are two-component spinors describing spin projections $\pm\nicefrac{1}{2}$
on the $z$ axis. We assume that the muon decays in the state $\phi_{+}$.
We will use the simplified notation $f,g=\tf,\tg$ and the dimensionless
variable $q=\frac{k}{m\alpha_{Z}}$,
\begin{align}
f & =\frac{2^{\gamma+1}\Gamma\left(\gamma+1\right)}{q\left(m\alpha_{Z}\right)^{\nicefrac{3}{2}}}\sqrt{\frac{\pi\left(1+\gamma\right)}{\Gamma\left(1+2\gamma\right)}}\text{Im}\left(1-iq\right)^{-\gamma-1},\\
g & =\frac{2^{\gamma+1}\left(1-\gamma\right)\Gamma\left(\gamma\right)}{\alpha_{Z}q^{2}\left(m\alpha_{Z}\right)^{\nicefrac{3}{2}}}\sqrt{\frac{\pi\left(1+\gamma\right)}{\Gamma\left(1+2\gamma\right)}}\text{Im}\left[1-iq\left(\gamma+1\right)\right]\left(1-iq\right)^{-\gamma-1},
\end{align}
where $\alpha_{Z}=Z\alpha$, $\gamma=\sqrt{1-\alpha_{Z}^{2}}$, and
$\alpha\simeq1/137$ is the fine structure constant. We employ the
basis (\cite{BjorkenQM}, eq. (3.7)), 
\begin{align}
w^{1}\left(\vk\right) & =c\left(\begin{array}{c}
1\\
0\\
\frac{k_{z}}{k^{0}+m}\\
\frac{k+}{k^{0}+m}
\end{array}\right),\quad w^{2}\left(\vk\right)=c\left(\begin{array}{c}
0\\
1\\
\frac{k_{-}}{k^{0}+m}\\
-\frac{k_{z}}{k^{0}+m}
\end{array}\right),\\
w^{3}\left(\vk\right) & =c\left(\begin{array}{c}
\frac{k_{z}}{k^{0}+m}\\
\frac{k_{+}}{k^{0}+m}\\
1\\
0
\end{array}\right),\quad w^{4}\left(\vk\right)=c\left(\begin{array}{c}
\frac{k_{-}}{k^{0}+m}\\
-\frac{k_{z}}{k^{0}+m}\\
0\\
1
\end{array}\right),
\end{align}
with $c=\sqrt{\frac{k^{0}+m}{2m}}$, $k^{0}=\sqrt{m^{2}+\vk^{2}}$,
and $k_{\pm}=k_{x}\pm ik_{y}$. In analogy with equation (A6) in \cite{Greub:1994fp},
we expand the bound wave function in this basis
\begin{align}
\Phi_{+}\left(\vk\right) &
                           =\sqrt{\frac{2m}{2k^{0}}}\left[
A_{+}w^{1}\left(\vk\right)
+A_{-}w^{2}\left(\vk\right)
+B_{+}^{\star}w^{4}\left(-\vk\right)
+B_{-}^{\star}w^{3}\left(-\vk\right)\right]\\
 & =\sqrt{\frac{k^{0}+m}{2k^{0}}}\left[A_{+}\left(\begin{array}{c}
1\\
0\\
\frac{k_{z}}{k^{0}+m}\\
\frac{k_{+}}{k^{0}+m}
\end{array}\right)+A_{-}\left(\begin{array}{c}
0\\
1\\
\frac{k_{-}}{k^{0}+m}\\
-\frac{k_{z}}{k^{0}+m}
\end{array}\right)
+B_{+}^{\star}
\left(\begin{array}{c}
-\frac{k_{-}}{k^{0}+m}\\
\frac{k_{z}}{k^{0}+m}\\
0\\
1
\end{array}\right)
+B_{-}^{\star}
\left(\begin{array}{c}
-\frac{k_{z}}{k^{0}+m}\\
-\frac{k_{+}}{k^{0}+m}\\
1\\
0 
\end{array}\right) 
\right].
\end{align}
For example, for the spin projection $+\nicefrac{1}{2}$, 
\begin{align}
A_{+} & =\sqrt{\frac{k^{0}+m}{2k^{0}}}\left(f+\frac{kg}{k^{0}+m}\right),\label{Alchemy:Aplus}\\
A_{-} & =0,\\
B_{+}^{\star} & =\sqrt{\frac{k^{0}+m}{2k^{0}}}k_{+}\left(-\frac{f}{k^{0}+m}+\frac{g}{k}\right),\\
B_{-}^{\star} & =\sqrt{\frac{k^{0}+m}{2k^{0}}}k_{z}\left(-\frac{f}{k^{0}+m}+\frac{g}{k}\right),
\end{align}
in agreement with (A7) in \cite{Greub:1994fp} except for $B_{-}^{\star}$,
for which we find the opposite overall sign. We proceed to check the
normalization,
\begin{equation}
\int\frac{\d^{3}k}{\left(2\pi\right)^{3}}\left(\left|A_{+}\right|^{2}+\left|A_{-}\right|^{2}+\left|B_{+}\right|^{2}+\left|B_{-}\right|^{2}\right)=1.\label{normalization}
\end{equation}
We confirm that the \textbf{$B_{\pm}$ }part of this integral is very
small, $0.16\%$ even for $Z=80$, in agreement with a comment below
(A8) in \cite{Greub:1994fp}. Indeed, the $B_{\pm}$ part of the normalization
integral, interpreted as the probability of finding a positron in
the atom, is $\order{\alpha_{Z}^{5}}$ when $Z\to0$. For this reason,
the negative energy components of the wave function were neglected
in \cite{Greub:1994fp}. The positive and negative energy components
can be separated by acting on the wave function with Casimir projectors,
\begin{align}
P_{A} & =\frac{m}{k^{0}}\left(w_{1}w_{1}^{\dagger}+w_{2}w_{2}^{\dagger}\right)=\frac{\slashed k+m}{2k^{0}}\gamma^{0},\\
P_{B} & =\gamma^{0}\frac{\slashed k-m}{2k^{0}},
\end{align}
such that $P_{A,B}^{2}=P_{A,B}$ and $P_{A}+P_{B}=1$. We find that
the rate calculated with $P_{A}$-projected wave functions is substantially
larger than when full wave functions are used. These results are compared
in Fig. \ref{fig:Rate-as-fnc-of-Z}, where we plot the rate divided
by
\begin{equation}
\Gamma_{0}=\frac{\GF^{2}m_{\mu}^{5}}{192\pi^{3}},
\end{equation}
the free muon decay rate at tree level, in the limit of a massless
electron; $\GF=\frac{\sqrt{2}g^{2}}{8M_{W}^{2}}$ is the Fermi constant
\cite{Marciano:1999ih}. The solid line Fig. \ref{fig:Rate-as-fnc-of-Z}
shows the full wave function result, and the dots -- the result with
projectors $P_{A}$. For small and moderate $Z$, up to $Z\simeq40$,
the results are close, and start to diverge quite strongly for larger
nuclei. 

Since the \textbf{$B_{\pm}$ }contribution to the normalization is
small even for large $Z$, these results are unexpected. We thus proceed
to check them in the position space. As a reward, we find that alternative
method to be simpler. It will allow us to derive a closed formula
for the rate in the limit of close electron and muon masses. 

\section{Bound $\mu^{-}$ to bound $e^{-}$ decay in position space\label{sec:R-space}}

In this section, we calculate the bound state transition rate $\left(Z\mu^{-}\right)\rightarrow\left(Ze^{-}\right)\nu_{\mu}\bar{\nu}_{e}$
using position space wave functions for the decaying muon and the
produced electron \cite{Berestetsky:1982aq},
\begin{align}
\Phi\left(\vr\right) & =\psi_{n=1,j=\frac{1}{2},\pm}\left(r,\theta,\phi\right)=\frac{f\left(r\right)}{\sqrt{4\pi}}u_{\pm},\label{eq:O1}
\end{align}
where 
\begin{equation}
f\left(r\right)=\left(2m\alpha_{Z}\right)^{3/2}\sqrt{\frac{1+\gamma}{2\Gamma\left(1+2\gamma\right)}}\left(2mr\alpha_{Z}\right)^{\gamma-1}\exp\left(-mr\alpha_{Z}\right),\label{eq:O2}
\end{equation}
and the mass $m$ is either $m_{\mu}$ for the muon or $m_{e}$ for
the electron. In the Dirac representation 
\begin{align}
u_{\pm} & =\cancel{\rho}\left(\begin{array}{c}
\phi_{\pm}\\
0
\end{array}\right),\label{eq:O3}
\end{align}
with $\rho^{\mu}=\left(\rho^{0},\vrho\right)=\left(1,i\frac{1-\gamma}{Z\alpha}\hat{r}\right)$.
Since this approach differs from Ref.~\cite{Greub:1994fp}, we present
it in some detail. 

\subsection{Factorizing neutrinos}

It is convenient to decompose the decay into two stages: first the
muon decays into the electron and a fictitious boson $A$; next, boson
$A$ decays into the $\nu\bar{\nu}$ pair. The kinematically allowed
range of the mass $m_{A}$ of the boson $A$, parametrized by a dimensionless
variable $z$, $m_{A}=zm_{\mu}$, is $z\in\left[0,z_{\max}=\left(E_{\mu}-E_{e}\right)/m_\mu\right]$
where $E_{\mu,e}$ are the muon and electron energies; they should
be replaced by muon and electron masses in the case of a free muon
decay. The decay rate is an integral over $z$,
\begin{equation}
\Gamma\left(\mu\to e\nu\bar{\nu}\right)=\frac{256\pi\Gamma_{0}}{g^{2}m_{\mu}}\int_{0}^{z_{\max}}\Gamma\left[\mu^{-}\to e^{-}A\right]z^{3}\d z,\label{factorGamma}
\end{equation}
where $g$ is the weak coupling constant. One advantage of eq. \eqref{factorGamma}
is that it holds both for a free and for a bound muon decay. It is
simpler to deal with a two-body decay $\mu\to eA$ than with $\mu\to e\nu\bar{\nu}$.
Binding effects as well as radiative corrections (ignored in the present
paper) affect only $\Gamma\left(\mu\to eA\right)$. 

As an example of using \eqref{factorGamma} consider a free muon decay.
Then $z_{\max}=1-\delta$, $\delta=\nicefrac{m_{e}}{m_{\mu}}$, and
\begin{equation}
\Gamma\left(\mu\to eA\right)=\frac{g^{2}}{32\pi}\left(1+z^{2}-2z^{4}+\delta^{2}z^{2}-2\delta^{2}+\delta^{4}\right)\frac{q\left(z\right)}{z^{2}},
\end{equation}
where $q\left(z\right)$ is the momentum of $A$; for a free muon
decay, $q\left(z\right)=\frac{\lambda^{\nicefrac{1}{2}}\left(1,\delta^{2},z^{2}\right)}{2}m_{\mu}$
with the Källén function $\lambda\left(x,y,z\right)=x^{2}+y^{2}+z^{2}-2\left(xy+yz+zx\right)$.
Integration over $z$ gives
\begin{equation}
\Gamma\left(\mu\to e\nu\bar{\nu}\right)=\Gamma_{0}\left(1-8\delta^{2}-24\delta^{4}\ln\delta+8\delta^{6}-\delta^{8}\right),\label{free}
\end{equation}
reproducing the correct electron mass dependence \cite{Pak:2008qt}.

\subsection{Decay rate\label{subsec:Decay-rate}}

The amplitude for the $\left(Z\mu\right)\to\left(Ze\right)+A$ transition
is
\begin{equation}
\calM=\frac{g}{\sqrt{2}}\int\d^{3}\vr\exp\left(i\vq\cdot\vr\right)\overline{\Phi}_{\mu}\left(\vr\right)\slashed\epsilon^{\lambda_{A}\star}L\Phi_{e}\left(\vr\right)\label{MrSpace}
\end{equation}
where $L=\frac{1-\gamma_{5}}{2}$ and $\lambda_{A}$ labels the polarization
state of $A$. The triple $\vr$ integration is done analytically.
Angular integrations lead to spherical Bessel functions, and the $r$-integration
results in a relatively compact formula. After squaring the amplitude,
we find, using $k_{A}=\sqrt{z_{\max}^{2}-z^{2}}$ and $a=\frac{1-\gamma}{\alpha_{Z}}$,
\begin{align}
\Gamma\left[\left(Z\mu^{-}\right)\rightarrow\left(Ze^{-}\right)+A\right] & =\frac{m_{\mu}g^{2}}{2\pi}k_{A}\left(N_{a}^{2}+N_{b}^{2}+F_{a}^{2}+F_{b}^{2}\right),\label{RateA}\\
N_{a} & =\sqrt{2}\frac{z_{\max}}{z}\left[4a^{2}\left(C_{2}-S_{3}\right)+\left(1+a^{2}\right)S_{1}\right],\nonumber \\
N_{b} & =\sqrt{2}\frac{k_{A}}{z}\left(1+a^{2}\right)S_{1},\nonumber \\
F_{a} & =4a^{2}\left(C_{2}-S_{3}\right)-2\left(1-a^{2}\right)S_{1},\nonumber \\
F_{b} & =4a\left(S_{2}-C_{1}\right),\nonumber 
\end{align}
where the quantities $C_{n}$ (and analogously $S_{n}$ with $\cos\to\sin$)
are
\begin{align}
C_{n} & =\frac{1+\gamma}{8}\left(\frac{4\delta}{\left(1+\delta\right)^{2}}\right)^{\gamma+\frac{1}{2}}\frac{\Gamma\left(1+2\gamma-n\right)}{k^{n}\Gamma\left(1+2\gamma\right)}\left(1+k^{2}\right)^{\frac{n-1}{2}-\gamma}\cos\left[\left(1+2\gamma-n\right)\arctan k\right],\\
k & =\frac{k_{A}}{\alpha_{Z}\left(1+\delta\right)}.
\end{align}
The rate $\Gamma\left[\left(Z\mu^{-}\right)\rightarrow\left(Ze^{-}\right)+\nu\bar{\nu}\right]$
can now be expressed as a single integral over $z$, a variable equivalent
to the invariant mass of the neutrinos, from zero to $z_{\max}=\gamma\left(1-\delta\right)$,
\begin{equation}
\frac{1}{\Gamma_{0}}\Gamma\left[\left(Z\mu^{-}\right)\rightarrow\left(Ze^{-}\right)+\nu\bar{\nu}\right]=128\int_{0}^{z_{\max}}\left(N_{a}^{2}+N_{b}^{2}+F_{a}^{2}+F_{b}^{2}\right)k_{A}z^{3}\d z.\label{rate-rSpace}
\end{equation}
This is the main result of this paper. Note that the position space
calculation results in a single integral representation for the rate.
This is much simpler than the result of Ref.~\cite{Greub:1994fp},
where two additional integration remain, over the magnitude and
the polar angle of the argument of the momentum space wave function.
Those integrations seem to be more involved than the corresponding
radial and angular integrations in the position space. In the following
section we perform the remaining integration over $z$ in the limit
of close electron and muon masses.

\section{Limiting Case of Similar Electron and Muon Masses\label{sec:EqualMasses}}

We expect out relusts to agree with Ref.~\cite{Greub:1994fp} for
small $Z$, since the only conceptual difference between our approaches
involves negative energy components, and those are suppressed by powers
of $\alpha_{Z}$. Here we demonstrate this agreement with a simple
closed formula in the limiting case of nearly equal masses. We write
\begin{equation}
m_{\mu}-m_{e}=\epsilon m_{\mu},\label{eq:3.1}
\end{equation}
and consider a hypothetical situation where $\epsilon$ is small.
In the limit $\epsilon\rightarrow0$, the decay rate computed in Ref.
\cite{Greub:1994fp} (cf. Eq. \eqref{eq:A7} in the Appendix) is
\begin{align}
\Gamma\left[\left(Z\mu^{-}\right)\rightarrow\left(Ze^{-}\right)\nu_{\mu}\bar{\nu}_{e}\right] & =\int_{0}^{m_{1}-m_{2}}d\left|\vq\right|\frac{G_{F}^{2}\vq^{2}}{12\pi^{3}}K\left(\left|\vq\right|\right)\nonumber \\
 & =\gamma^{5}\epsilon^{5}\frac{G_{F}^{2}m_{\mu}^{5}}{15\pi^{3}}\left\{ \left[F_{1}\left(0\right)-F_{2}\left(0\right)\right]^{2}+F_{1}\left(0\right)F_{2}\left(0\right)\right\} .\label{eq:3.2}
\end{align}
In this approximation of equal masses of muon and electron, the momentum
transferred to the neutrinos $\mathbf{q}$ is approximately zero.
Therefore, the form factors $F_{1}$ and $F_{2}$ given in Eq. \eqref{eq:3.2}
are evaluated at $\left|q\right|=0$ which gives 
\begin{align}
F_{1}\left(0\right) & =\int\frac{d^{3}k}{\left(2\pi\right)^{3}}\psi_{\mu}\left(\vk\right)\psi_{e}^{*}\left(\vk\right)\frac{2k^{0}+m_{\mu}}{3k^{0}}=\frac{2}{3}+\frac{1}{3}\left\langle L^{-1}\right\rangle ,\nonumber \\
F_{2}\left(0\right) & =\int\frac{d^{3}k}{\left(2\pi\right)^{3}}\psi_{\mu}\left(\vk\right)\psi_{e}^{*}\left(\vk\right)\frac{k^{0}-m_{\mu}}{3k^{0}}=\frac{1}{3}-\frac{1}{3}\left\langle L^{-1}\right\rangle ,\label{eq:3.3}
\end{align}
where the mean inverse Lorentz factor $\left\langle L^{-1}\right\rangle $
is 
\[
\left\langle L^{-1}\right\rangle =\int\frac{d^{3}k}{\left(2\pi\right)^{3}}\left|\psi_{\mu}\left(\vk\right)\right|^{2}\frac{m_{\mu}}{k_{1}^{0}}.
\]
Hence, Eq. \eqref{eq:3.2} becomes 
\begin{equation}
\frac{\Gamma}{\Gamma_{0}}=\frac{64}{5}\epsilon^{5}\gamma^{5}\frac{1+\left\langle L^{-1}\right\rangle +\left\langle L^{-1}\right\rangle ^{2}}{3}.\label{eq:3.4}
\end{equation}
The numerical results $\frac{\Gamma\left[\left(Z\mu^{-}\right)\rightarrow\left(Ze^{-}\right)\nu_{\mu}\bar{\nu}_{e}\right]}{\Gamma_{0}}$
for the limiting case of almost equal masses are stated in the Table
\ref{tab1.3}.

\begin{table}
\centering{}%
\begin{tabular}{cccc}
\hline 
\hline 
$Z$ & Eq. \eqref{eq:A6} & Eq. \eqref{eq:3.4} & Eq. \eqref{eq:3.5}\tabularnewline
\hline 
10 & $1.25\cdot10^{-9}$ & $1.26\cdot10^{-9}$ & $1.26\cdot10^{-9}$\tabularnewline
80 & $3.85\cdot10^{-10}$ & $3.83\cdot10^{-10}$ & $3.72\cdot10^{-10}$\tabularnewline
\hline 
\hline 
\end{tabular}\caption{\label{tab1.3} Numerical values of
  $\Gamma\left[\left(Z\mu^{-}\right)\rightarrow\left(Ze^{-}\right)\nu_{\mu}\bar{\nu}_{e}\right] /
{\Gamma_{0}}$
for $\epsilon=0.01$ for a small $Z=10$ and a large $Z=80$: using
the formalism of Ref.~\cite{Greub:1994fp}, Eq.~\eqref{eq:A6} (second
column), its limit for $m_{e}\simeq m_{\mu},$ Eq.~\eqref{eq:3.4}
(third column), and the $m_{e}\simeq m_{\mu}$ limit of our approach,
Eq.~\eqref{eq:3.5} (forth column). As expected, the agreement is
better for small $Z$ (first line). }
\end{table}

In equal mass limit, using $\left|\vq\right|r\to0$, our momentum
space as well as position space treatments lead to the expression
\begin{equation}
\frac{\Gamma\left[\left(Z\mu^{-}\right)\rightarrow\left(Ze^{-}\right)\nu_{\mu}\bar{\nu}_{e}\right]}{\Gamma_{0}}=\frac{64}{5}\epsilon^{5}\gamma^{5}\frac{1+\gamma+\gamma^{2}}{3}.\label{eq:3.5}
\end{equation}
The analogous limit of the free muon decay rate, eq. \eqref{free},
is $\frac{64}{5}\epsilon^{5}$. Thus the binding effects given by
$\gamma^{5}\frac{1+\gamma+\gamma^{2}}{3}=1-3\alpha_{Z}^{2}+\order{\alpha_{Z}^{4}}$
in case of the decay into a bound electron are more pronounced than
in the case of the decay into a free electron, given in eq. \eqref{eq:TimeDil}.
For a free electron, the effect can be approximated by a single factor
of $\gamma\simeq1-\alpha_{Z}^{2}+\order{\alpha_{Z}^{4}}$. 

Numerical evaluation of Eq. \eqref{eq:3.5} is given in the last column
of Table \ref{tab1.3}. For small $Z$, we have $\left\langle L^{-1}\right\rangle \approx\gamma$
and hence the corresponding results coincide in this case of equal
muon and electron masses. This is not the case for large $Z$, where
$\left\langle L^{-1}\right\rangle $ is larger then $\gamma$.

\section{Conclusion\label{sec:Conclusion}}

We have calculated the decay rate of a bound muon to bound electron
using Dirac wave functions for different values of $Z$ in two formalisms.
Numerical results in momentum and in position space coincide, provided
that complete wave functions (both positive and negative energy components)
are used. If the negative energy parts of the wave functions are neglected,
as was done in Ref.~\cite{Greub:1994fp}, the results are significantly
larger. For $Z=80$ the difference is about $38\%$. 

This is surprising since the probability of finding positrons in a
hydrogen-like atom or ion is very small even for $Z=80$. Our tentative
interpretation is that the probability of the decay into a bound electron
is very suppressed and that this suppression is relatively less severe
for the negative energy components. We note that the decay vertex
couples positive and negative energy components without a suppression
factor of $\alpha_{Z}$. Also, the decay rate involves an interference
of large $A_{\pm}$ with small $B_{\pm}$ wave function terms, whereas
the normalization integral involves the small $B_{\pm}$ only in second
powers, thus greatly decreasing their contribution (see eq. \eqref{normalization}). 

In order to check this unexpected result, we developed a position
space approach. It resulted in a simple one-dimensional integral representation
of the rate, eq. \eqref{rate-rSpace}. The remaining integral has
been done in the limiting case of close electron and muon masses,
eq. \eqref{eq:3.5}, giving a closed formula valid for all $\alpha_{Z}$.

In closing, we quote from Sidney Coleman's field theory lectures \cite{Chen:2018cts}:
\emph{Dirac\textquoteright s theory gives excellent results to order
$\left(\nicefrac{v}{c}\right)^{2}$ for the hydrogen atom, even without
considering pair production and multi-particle intermediate states.
This is a fluke. }

\section*{Acknowledgement}

A.C.~thanks Andrey Volotka for conversations that stimulated this
project.  M.J.A. and A.C. gratefully acknowledge the hospitality of
the Banff International Research Station (BIRS) where parts of this
work were done.  This research was supported by the Natural Sciences
and Engineering Research Council of Canada.

\appendix

\section{Bound Muon Decay in the Formalism of Ref.~\cite{Greub:1994fp}\label{sec:Alchemy}}

In this Appendix, we summarize the formalism of Ref.~\cite{Greub:1994fp}
for the transition $B_{1}\to B_{2}+X$, where $B_{1}$ and $B_{2}$
are bound states.

The invariant amplitude of $\left(Z\mu^{-}\right)\rightarrow\left(Ze^{-}\right)\nu_{\mu}\bar{\nu}_{e}$
decay is 
\begin{equation}
\mathcal{M}_{B_{1}\rightarrow B_{2}}=\frac{4G_{F}}{\sqrt{2}}\sqrt{4m_{B_{1}}m_{B_{2}}}N_{\mu}S_{sr}^{\mu},\label{eq:A1}
\end{equation}
where the subscripts $B_{1}$ and $B_{2}$ represent the $\left(Z\mu^{-}\right)$
and $\left(Ze^{-}\right)$ states, respectively. The masses of the
bound states are, with $M$ denoting the nucleus mass (note that in
our approach the nucleus is treated as infinitely heavy and $M$ does
not appear),
\begin{equation}
\begin{array}{c}
m_{B_{1}}=M+m_{1},\:m_{1}=m_{\mu}-E_{b,\mu},\\
m_{B_{2}}=M+m_{2},\:m_{2}=m_{e}-E_{b,e},
\end{array}\label{eq:A2}
\end{equation}
where $E_{b,\mu\left(e\right)}$ are the binding energies. In Eq.
(\ref{eq:A1}), the neutrino part is given by 
\begin{equation}
N_{\mu}=\bar{u}\left(p_{\nu_{\mu}}\right)\gamma_{\mu}L\upsilon\left(p_{\nu_{e}}\right),\label{eq:A3}
\end{equation}
and the charged current part is 
\begin{equation}
S_{sr}^{\mu}=\int\frac{\d^{3}k_{1}}{\left(2\pi\right)^{3}}\psi_{\mu}\left(\vk_{1}\right)\psi_{e}^{*}\left(\vk_{1}-\vq\right)\frac{\bar{u}_{s}\left(e;\vk_{1}-\vq\right)}{\sqrt{2k_{2}^{0}}}\gamma^{\mu}L\frac{u_{r}\left(\mu;\vk_{1}\right)}{\sqrt{2k_{1}^{0}}}.\label{eq:A4}
\end{equation}
Here, $k_{1}$, $k_{2}$, $p_{\nu_{e}}$ and $p_{\nu_{\mu}}$ are
the 4-momenta of the muon, electron, $\nu_{e}$ and $\nu_{\mu}$,
respectively, and the subscripts $r$ and $s$ are spin indices $\left(r,s=\pm1/2\right)$.

The corresponding decay rate for $\left(Z\mu^{-}\right)\rightarrow\left(Ze^{-}\right)\nu_{\mu}\bar{\nu}_{e}$
can be calculated as 
\begin{equation}
\d\Gamma=\frac{1}{2m_{B_{1}}}\d\Phi\left|\mathcal{M}_{B_{1}\rightarrow B_{2}}\right|^{2},\label{eq:A5}
\end{equation}
After integration over the phase space and neglecting terms suppressed
by $1/M$, the decay rate is \cite{Greub:1994fp} 
\begin{align}
\Gamma & =\frac{G_{F}^{2}}{12\pi^{3}}\int_{0}^{m_{1}-m_{2}}\d\left|\vq\right|\vq^{2}K\left(\left|\vq\right|\right)\label{eq:A6}
\end{align}
where 
\begin{align}
K\left(\left|\vq\right|\right) & =\left[q^{2}+2\left(m_{1}-m_{2}\right)^{2}\right]\left(F_{1}^{2}+F_{2}^{2}\right)+\frac{q^{2}}{m_{\mu}^{2}}\left[4\left(m_{1}-m_{2}\right)^{2}-q^{2}\right]\left(F_{3}^{2}+F_{4}^{2}\right)\nonumber \\
 & -6q^{2}\left[F_{1}F_{2}+\frac{q^{2}}{m_{\mu}^{2}}F_{3}F_{4}+\frac{m_{1}-m_{2}}{m_{\mu}}\left(F_{1}-F_{2}\right)\left(F_{3}-F_{4}\right)\right],\label{eq:A7}
\end{align}
and $q^{2}=q_{0}^{2}-\vq^{2}=\left(m_{1}-m_{2}\right)^{2}-\vq^{2}.$
The form factors $F_{i}$ are defined as
\begin{equation}
F_{i}\left(q^{2}\right)=\int\frac{d^{3}k_{1}}{\left(2\pi\right)^{3}}\psi_{\mu}\left(\vk_{1}\right)\psi_{e}^{*}\left(\vk_{1}-\vq\right)\frac{h_{i}}{\sqrt{4k_{1}^{0}k_{2}^{0}\left(k_{1}^{0}+m_{\mu}\right)\left(k_{2}^{0}+m_{e}\right)}},\label{eq:A8}
\end{equation}
with 
\begin{align}
h_{1} & =\left(k_{1}^{0}+m_{\mu}\right)\left(k_{2}^{0}+m_{e}\right)+q^{0}\left[\left(1-C\right)\left(k_{1}^{0}+m_{\mu}\right)-C\left(k_{2}^{0}+m_{e}\right)\right]+\left(B-C\right)q_{0}^{2}+A,\nonumber \\
h_{2} & =\left(C-B\right)q^{2}+2A,\nonumber \\
h_{3} & =\left[\left(1-C\right)\left(k_{1}^{0}+m_{\mu}\right)+\left(B-C\right)q_{0}\right]m_{\mu},\label{eq:A9}\\
h_{4} & =\left[C\left(k_{2}^{0}+m_{e}\right)-\left(B-C\right)q_{0}\right]m_{\mu}\nonumber 
\end{align}
and the expressions of $A,B$ and $C$ are given in Eq. (35) of \cite{Greub:1994fp}.
We mention that in the expressions for $h_{1}$ and $h_{2}$ in Eq.
\eqref{eq:A9}, the sign of $A$ is different than in \cite{Greub:1994fp}
(cf. Eq. (34) there).


\end{document}